\documentclass{vgtc}                
\makeatletter
\renewcommand*{\@fnsymbol}[1]{\@arabic{#1}}

\ifpdf
\pdfoutput=1\relax                   
\usepackage{graphicx}                
\DeclareGraphicsExtensions{.pdf,.png,.jpg,.jpeg} 
\else
\usepackage{graphicx}                
\DeclareGraphicsExtensions{.eps}     
\fi%

\graphicspath{{figures/}{pictures/}{images/}{./}} 

\usepackage{microtype}                 
\PassOptionsToPackage{warn}{textcomp}  
\usepackage{textcomp}                  
\usepackage{mathptmx}                  
\usepackage{times}                     
\usepackage{cite}                      
\usepackage{tabu}                      
\usepackage{booktabs}                  

\usepackage[bookmarks=false]{hyperref}
\usepackage[]{quoting}
\usepackage{multirow}

\usepackage{fancyhdr}

\usepackage{tikz}

\usepackage{tabularx}
\usepackage{booktabs}
\usepackage{ctable}
\usepackage{array}
\usepackage{multirow}
\usepackage{colortbl}

\newcolumntype{L}{>{\raggedright\arraybackslash}X}
\newcolumntype{C}{>{\centering\arraybackslash}X}
\newcolumntype{R}{>{\raggedleft\arraybackslash}X} 
\definecolor{c1}{rgb}{0.95,0.95,0.95}
\definecolor{c2}{rgb}{0.88,0.88,0.88}
\definecolor{c3}{rgb}{0.8,0.8,0.8}
\newcommand{\cca}{\cellcolor{c1}}
\newcommand{\ccb}{\cellcolor{c2}}
\newcommand{\ccc}{\cellcolor{c3}}
\usepackage{anyfontsize}

\usepackage[utf8]{inputenc}

\onlineid{0}

\vgtccategory{Research}
\vgtcpapertype{please specify}

\title{How Visualization PhD Students Cope with Paper Rejections}

\author{Shivam Agarwal$^\dag$\thanks{shivam.agarwal@paluno.uni-due.de}, Shahid Latif$^\dag$\thanks{shahid.latif@paluno.uni-due.de}, and Fabian Beck\thanks{fabian.beck@paluno.uni-due.de}
}

\affiliation{\scriptsize $^\dag$ Equal contribution \\paluno, The Ruhr Institute of Software Technology \\University of Duisburg-Essen, Germany}

\abstract{
	We conducted a questionnaire study aimed towards PhD students in the field of visualization research to understand how they cope with paper rejections. We collected responses from 24 participants and performed a qualitative analysis of the data in relation to the provided support by collaborators, resubmission strategies, handling multiple rejects, and personal impression of the reviews. The results indicate that the PhD students in the visualization community generally cope well with the negative reviews and, with experience, learn how to act accordingly to improve and resubmit their work. Our results reveal the main coping strategies that can be applied for constructively handling rejected visualization papers. The most prominent strategies include: discussing reviews with collaborators and making a resubmission plan, doing a major revision to improve the work, shortening the work, and seeing rejection as a positive learning experience. 
} 

\vgtcinsertpkg

\newcommand\copyrightnotice{%
	\begin{tikzpicture}[remember picture,overlay]
	\node[anchor=north,yshift=0pt] at (current page.north) {\parbox{\dimexpr\textwidth-\fboxsep-\fboxrule\relax}{ \begin{center}
					\textit{To appear in Celebrating the Scientific Value of Failure (FailFest) Workshop at IEEE VIS 2020.}
				\end{center}}};
	\end{tikzpicture}%
}

\begin{document}


\firstsection{Introduction}
\maketitle

\copyrightnotice

Facing paper rejections in academia is common and sometimes has adverse effects on individuals. Existing research studies the effect of paper rejections on scholars~\cite{sierk2016the, edwards2018emotions} and provide some guidelines on how to cope with such experiences (\textit{e.g.}, ~\cite{day2011thesilent, rubin2014converting}). However, the studies were conducted for management and statistical sciences and the results might not necessarily generalize to the visualization community due to the diversity in reviewing models and culture. Additionally, these studies did not focus on PhD students specifically. We are interested in this group because we want to understand, among other things, if and how PhD students are negatively affected by paper rejections as well as what are their strategies to cope with it. PhD students, as new and inexperienced members of the community, might suffer more from negative reviews and paper rejections. If badly timed, a single rejected paper can lead to significant delays in completing the PhD dissertation or even threaten its successful completion.
%
%
%

Within visualization research, there exists literature that outlines common reasons of a paper rejection and provides guidelines for preparing a manuscript for submission~\cite{munzner2008process}. Additionally, senior researchers of the visualization community have made suggestions on how to deal with paper rejections. Elmqvist~\cite{elmqvist2016dealing,elmqvist2018above} described some strategies on how to handle paper rejections while acknowledging that young PhD students are most vulnerable. Taking a wider perspective, Shneiderman~\cite{shneiderman2016new} laid his thoughts on a new paradigm of research providing guidance to the students, junior and senior researchers, and policymakers. However, we lack research that studies how PhD students in the visualization research community perceive and handle paper rejections. 

In this paper, we aim at understanding and revealing the strategies of visualization PhD students in coping with paper rejections. We conducted an online study to gather responses from our target user group, \textit{i.e.}, PhD students working in visualization research community. Based on the responses, we performed a qualitative data analysis and report our findings. 
We provide our questionnaire, raw data, and coded responses as a supplemental material~\cite{supplementary}. 

\section{Study Design}

We decided to use a short online questionnaire for the study to reach visualization PhD students internationally. To not bias the participants towards certain answers and to also discover unexpected strategies, the questionnaire used open-ended questions for addressing the main research questions. Hence, a mostly qualitative analysis is required to identify common themes and noteworthy exceptions. To capture the full breadth of the community, we strove towards diversity of participants with respect to the country of affiliation and level of experience (among PhD students).


\subsection{Questionnaire}
The online questionnaire consists of three sections: \textit{general information}, \textit{publication history}, and \textit{handling paper rejections}. The questions in the first section focus on knowing the stage of participants' doctoral studies, countries of their universities, and field of study during their bachelor and master studies. The second section has two questions and aims at gauging the research experience of participants in terms of submitted and rejected number of peer-reviewed research articles. The third and final section captures the experience of participants with the papers that were rejected and re-submitted. All questions in the third section were open-ended questions and participants could write as much as they wanted. In particular, we asked the following four questions:

\begin{itemize}
\item \textbf{Q1:} How did your lab-mates/colleagues/supervisors help to handle paper rejections? 
\item \textbf{Q2:} What strategic decisions did you take to re-submit the rejected papers?
\item \textbf{Q3:} How did you handle the papers that got rejected more than once? (if any)
\item \textbf{Q4:} How did your impression of paper reviews change over time? 
\end{itemize}

These questions focus on capturing the possible ways or steps that students may have taken to cope when they receive a paper rejection.
The questionnaire was designed to be completed in approximately 25--30 minutes.

\subsection{Target Participants and Recruiting}
The target participants were the PhD students from the visualization community who have submitted research articles to peer-review venues and have already experienced one or more paper rejections. This target group ranges from first year PhD students to the ones who have recently (in the past 6 months) graduated or left the PhD program. The questionnaire was posted to the IEEE VIS and Infovis.org mailing lists as well as to the Discord channel used by EuroVis 2020 and Euro Graphics 2020. We also forwarded the questionnaire within our professional network by email and advertised it on Twitter. 

In total, we received 43 responses until the submission date. In this paper, as responses have been still coming in during data analysis, we focus on the first 24 responses and plan on including the later responses in an extended version of this paper. The following analysis and descriptions are purely based on the first 24 responses. 




\subsection{Analysis Methodology}
To analyze the responses, we followed an open coding process. The process began with two of the authors independently coding the first five responses, followed by a discussion on a merged set of authoritative codes. This resulted in a list of 19 codes. In the next stage, we divided the remaining responses into two equal parts (9 and 10 responses in these groups) and each coder used the list of already established 19 codes to do the coding. During this second stage of coding, both coders still came across new types of comments and occasionally had to define new codes. These codes were then immediately added to the shared list of codes. Notifying the other coder allowed him to use the newly defined code as well. This process, once again, resulted in some partly redundant codes that were again merged and resolved in another follow-up meeting. This refinement left us with a set of 28 different codes. Now, both coders used this set of codes to validate each other's coding. The resulting conflicts during the validation phase were then resolved in subsequent meetings, which also led to a further consolidation to 25 codes (cf. \autoref{tab:frequency}). 


\begin{figure}[tb]
    \centering
    \includegraphics[width=1.0\columnwidth,]{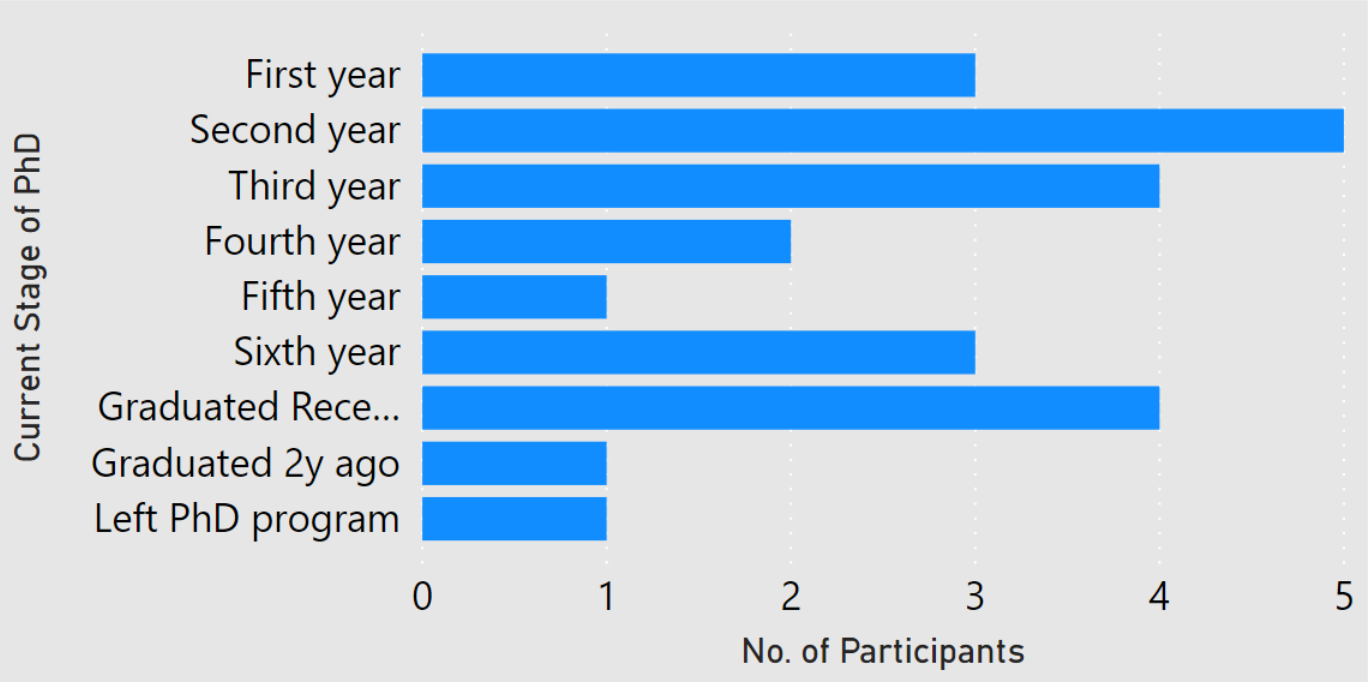}
   \caption{Stage of PhD studies of the participants. 
   }
   \label{fig:phdstage}
\end{figure}

\begin{figure}[t]
    \centering
      \includegraphics[width=1.0\columnwidth,]{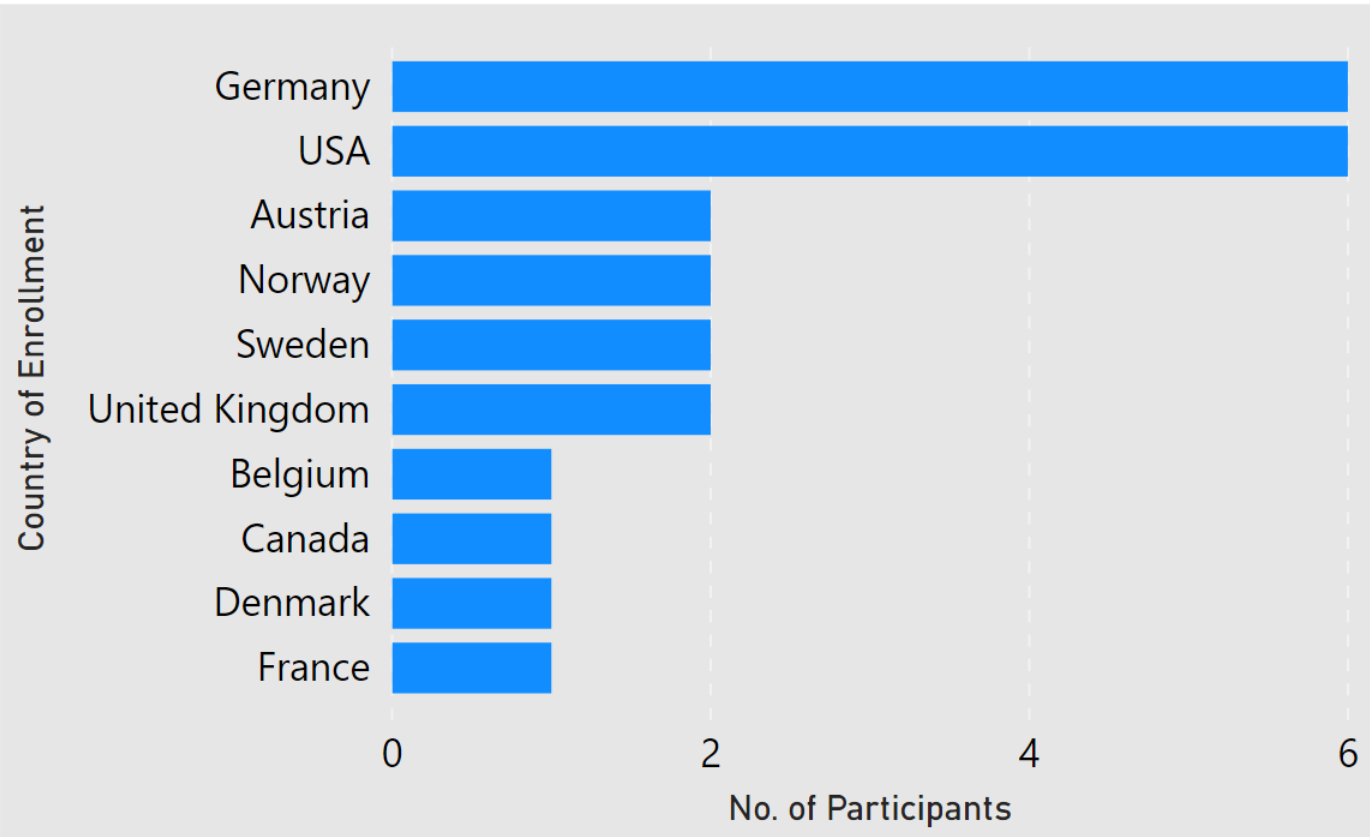}
   \caption{Participants by country of affiliation. 
   }
   \label{fig:country}
\end{figure}

\begin{figure}[t]
    \centering
      \includegraphics[width=1.0\columnwidth, trim = {0.0in 0.0in 0.8in 0.0in}, clip=true ]{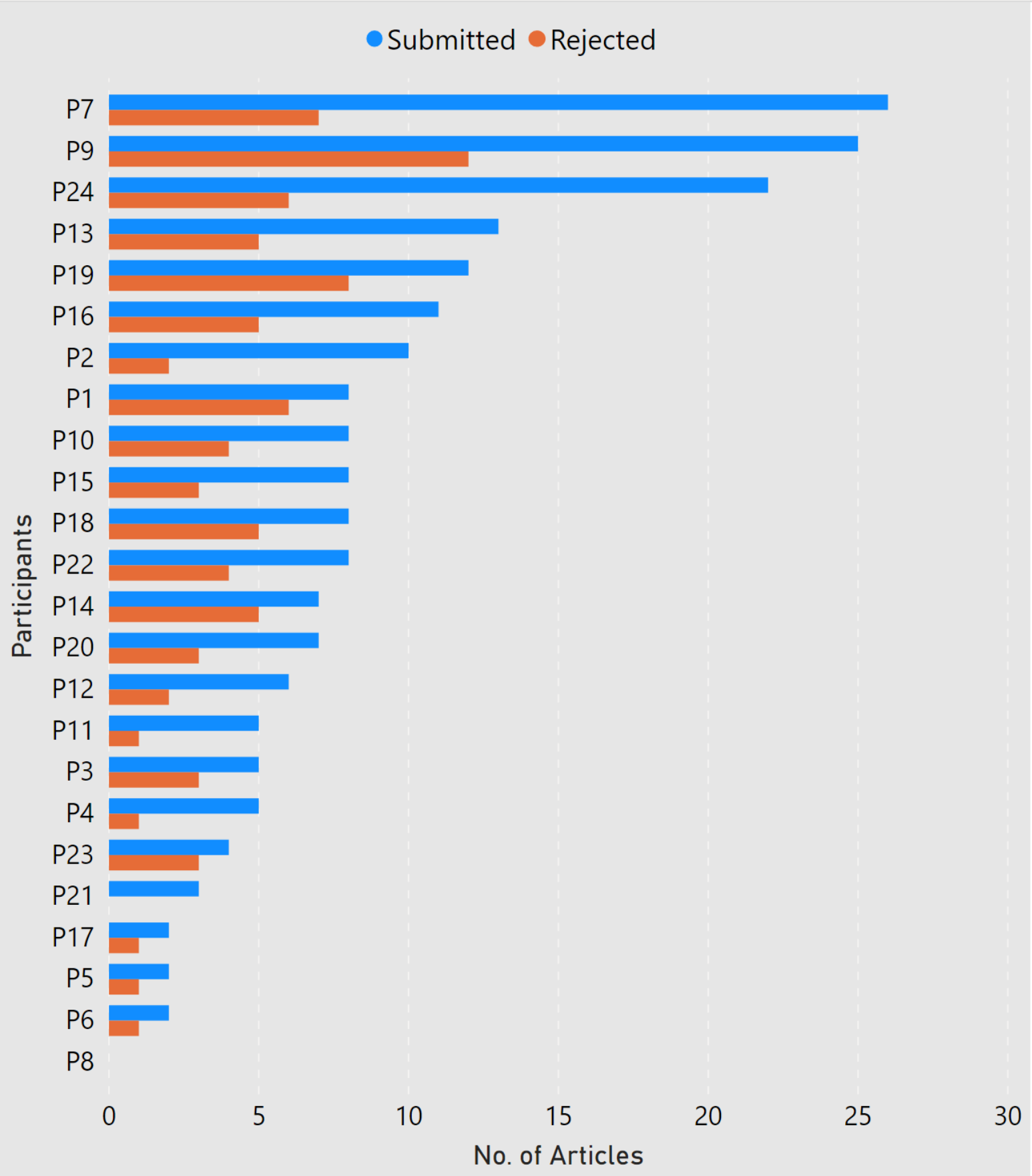}
   \caption{Number of submitted and rejected articles per participant. 
   }
   \label{fig:subrej}
\end{figure}

\section{Results}

Based on the assigned codes (\autoref{tab:frequency}), we report the results per question (Q1--Q4), each discussing the most frequently mentioned types of answers and noteworthy individual examples. We begin with summarizing the \textit{general information} and \textit{publication history} of the participants. We numbered the participants with identifiers P1--P24. Individual participants are listed when referring to groups of five participants or fewer; else, we just provide the count of participants, but details can be found in the supplemental material~\cite{supplementary}. 
\begin{table}[tb]
\centering
\caption{List of codes along with their frequency of occurrence per question and total (sorted by total).}
\label{tab:frequency}
\fontsize{7}{8}\selectfont
\sffamily
\begin{tabularx}{\columnwidth}{lCCCCC}
\toprule
\multicolumn{1}{c}{\multirow{2}{*}{\textbf{Codes}}}                       & \multicolumn{5}{c}{\textbf{Frequency}}                                                     \\ \cline{2-6}
\multicolumn{1}{c}{}                                                      & \textbf{Q1} & \textbf{Q2} & \textbf{Q3} & \textbf{Q4} & \textbf{Total} \\ \midrule
improvement                                                                 & \ccb 4      & \ccc 13                          & \ccc 11      & \ccb 4      & \ccc 32        \\ 
downgrade work                                                              & --          & \ccc 16                          & \ccb 5      & --          & \ccc 21        \\ 
discuss reviews                                                             & \ccc 19     & --                               & \cca 1      & --          & \ccc 20        \\ 
impression gets better                                           & --          & --                               & --          & \ccc 12     & \ccc 12        \\ 
see rejection as positive                                                               & --          & \cca 1                           & \ccb 4      & \ccb 5      & \ccc 10        \\ \midrule
resubmission plan                                                           & \ccb 9      & --                               & --          & --          & \ccb 9         \\ 
similar venue scope                                                         & --          & \ccb 9                           & --          & --          & \ccb 9         \\ 
painful reviews at first                                                    & --          & --                               & --          & \ccb 9      & \ccb 9         \\ 
informal discussion                                                         & \ccb 6      & --                               & --          & \cca 2      & \ccb 8         \\ 
impression remains the same                                                 & --          & --                               & --          & \ccb 7      & \ccb 7         \\ \midrule
submit to journal                                                           & --          & \ccb 6                           & --          & --          & \ccb 6         \\ 
fair reviews                                                                & \cca 1      & --                               & --          & \ccb 5      & \ccb 6         \\ 
high-level reflection of reviews                                            & \cca 3      & --                               & --          & \cca 2      & \ccb 5         \\ 
discuss reviews quickly                                                     & \cca 3      & --                               & --          & --          & \cca 3         \\ 
contradictory reviews                                                       & --          & --                               & --          & \cca 3      & \cca 3         \\ \midrule
different venue scope                                                       & --          & \cca 2                           & --          & --          & \cca 2         \\ 
disagree with reviews                                                       & --          & --                               & --          & \cca 2      & \cca 2         \\ 
no special discussion                                                       & \cca 1      & --                               & --          & --          & \cca 1         \\ 
lack of support from coauthors                                              & \cca 1      & --                               & --          & --          & \cca 1         \\ 
encouragement from supervisor                                               & \cca 1      & --                               & --          & --          & \cca 1         \\ \midrule
dealing with  rejections meeting                                            & \cca 1      & --                               & --          & --          & \cca 1         \\ 
supervisor's decision                                                       & --          & \cca 1                           & --          & --          & \cca 1         \\ 
post on arxiv                                                               & --          & \cca 1                           & --          & --          & \cca 1         \\ 
unclear reviews                                                             & --          & --                               & --          & \cca 1      & \cca 1         \\ 
get early feedback                                                          & --          & \cca 1                           & --          & --          & \cca 1             \\ \midrule
\textbf{Total}                                                              & \textbf{49} & \textbf{50}                      & \textbf{21} & \textbf{52} & \textbf{172}   \\ \bottomrule
\end{tabularx}
\end{table}

\subsection{Participants}

Among the 24 participants, about 84 percent of the participants had done a master degree in computer science, and almost 63 percent had completed a bachelors in computer science. As shown in \autoref{fig:phdstage}, a majority of the participants (22) were PhD students ranging from first year to sixth year and the ones who graduated in the past six months. One respondent (P2) had left the PhD program and one participant (P22) had graduated two years ago. 
Participants were enrolled at universities in 10 different countries, with half of the responses coming from Germany and USA (~\autoref{fig:country}). None of the participant claimed of being enrolled with multiple universities. Among the participants, the number of submitted and rejected peer-reviewed articles ranged from 0 to 26 and 0 to 12 respectively (\autoref{fig:subrej}). One participant (P8, a second year student) did not submit any article to date. Another participant (P21, a first year student) had not experienced any rejection. But since both of them provided valuable comments in their responses, we decided to still include them in our analysis.





\subsection{Support of Collaborators (Q1)}

As seen from ~\autoref{tab:frequency}, the most mentioned support strategies within the collaboration groups of participants were to \textit{discuss reviews} (19 participants) and make a concrete \textit{resubmission plan} (9 participants) with collaborators of the paper. Six participants reported to have had \textit{informal discussions} (\textit{e.g.}, \textit{vis-à-vis} during a coffee or lunch break, or via instant messaging) with their colleagues. They either \textit{discussed the reviews quickly} (P3, P9, and P14) or celebrated the success of their peers and commiserated with the others who faced a rejection (P15 and P24). 
For getting support within the work group, the similarity of research topics seems to play a role as P14 commented: \textit{``I just talked briefly with my supervisor about it. Since I'm the only person in my lab doing visualization research, the other ones don't know about the problems so much in detail.''} 

Towards improving the submission, participant P7 complained about the \textit{lack of support from coauthors} in revising or reworking the paper and mentioned that they did not have background in visualization research. 
Four participants (P2, P7, P10, and P12) discussed the exact \textit{improvements} and feasibility of these improvements for their paper with coauthors. Three experienced students (P10, P13, and P19) stated that they did a \textit{high-level reflection of reviews} to understand what went wrong and to find areas of improvement.



\subsection{Strategies for Resubmission (Q2)}

When it comes to resubmitting rejected papers, prominent strategies are either \textit{downgrading work} to resubmit in a lower-ranked venue (16 participants) or making \textit{improvements} in the work (13 participants). The \textit{improvement} might involve considerably extending the work with additional contributions (P12) or by including a user study (P1, P3, and P6). Both P1 and P12 stated that they  converted short papers to full papers.
The participants who decided to go for an \textit{improvement} either resubmit their papers to a \textit{similar scoped venue} (9 participants) or to a \textit{different scoped venue} (P7 and P15). It could be better to switch to a different venue for better aligning contributions with the target audience as P15 stated: \textit{``In the case of one paper the reviewers commented that the novelty/contribution to the visualization community was limited, so here we submitted the application-based paper to a workshop with a better audience fit [\ldots].''} 


Being afraid that somebody else might publish similar research first, participant P24 mentioned that they had \textit{posted} the paper on \textit{arXiv} and described it as: \textit{``[\ldots] if we were scared of being `scooped' we would put it on arXiv and go for the next venue. If no external pressure, we hold on to next year's visualization-related conference.''}

The responses also indicated few \textit{submission} strategies of the papers. Participant P7 reported his/her tendency of submitting a poster to \textit{get early feedback}: \textit{``I tend to like to do posters, because I can try to think about how to present the work visually, and in verbal discussions with others.  This is generally helpful for me.''} Another strategy is to \textit{submit to a journal} (6 participants) rather than waiting for the next conference cycle. P4 responded that he/she prefers to directly submit to journals as it provides more review cycles (major and/or minor revisions) in contrast to conferences where a major revision is considered as a reject. Participant P14 also favored this strategy as some journals in visualization community (\textit{e.g.}, TVCG) offers the presentation of accepted work at any major visualization conference. 

\subsection{Handling Multiple Rejects (Q3) }
Eleven participants mentioned that they had made further \textit{improvements} before resubmitting the papers that were rejected more than once.
Similar to handling first rejections of their papers, the participants reported to \textit{downgrade work} by resubmitting to lower-rated venues (P7, P15, P19, P20, and P23). None of the participants mentioned to have abandoned their work after getting it rejected multiple times. However, participant P7 highlighted that shortening the work could be a good strategy to avoid spending more time on a project. Four participants (P7, P14, P15, and P20) mentioned that they had \textit{seen rejection as a positive} learning experience while handling papers that got multiple rejects. The response of participant P7 summarizes this positive outlook:
\begin{quoting}
\textit{``My attitude is `this will get in somewhere, eventually'.  I have seen my papers improve dramatically after revising and resubmitting, so I tend to think of rejections as a good thing, because the paper wasn't ready and will only get better each time I try again. If it were to be on a topic where I didn't feel like making substantial changes, I might stop at submitting a short paper or a poster, but I haven't encountered this yet.''} -- P7
\end{quoting}
However, participant P14 also highlighted the difficulty while handling papers that got multiple rejections as: \textit{``It was really hard to not give up. There were many tears. But in the end, you have to stand up again and improve your work.''}. Eleven participants either did not provide any answer to the question or mentioned that, until now, they had not faced multiple rejections of the same paper.

\subsection{Impression of Rejected Paper Reviews (Q4)}

Nine participants highlighted that it was painful to read the reviews at first, especially, on the notification day. Considering how the subjective impression of the reviews changes over time, half of the participants (12) mentioned that their  \textit{impression got better}, while seven participants reported that their \textit{impression remained the same}. 
Four participants (P5, P6, P7, and P15) said that they \textit{saw rejection as positive}. A response from participant P6 summarizes a strategy to avoid taking reviews personally: ``\textit{[The impression of the reviews] fairly remained the same. I always took it constructively and know that rejections are part of the science process. It’s not [a] reflection on me personally or my skills.}''.

Five participants (P3, P12, P14, P20, and P22) reported that overall they had gotten \textit{fair reviews} while two of them (P3 and P22) \textit{disagreed with reviews} on some points. Three participants (P1, P7, and P14) reported that sometimes reviewers had provided \textit{contradictory reviews}, while participant P18 mentioned receiving \textit{unclear reviews}. Participant P13 mentioned a strategy of not considering each point in a review; instead he/she had learned to do a \textit{high-level reflection of reviews} with experience:
\begin{quoting}
\textit{``My perception of specific reviews did not change much with time, but my general view on reviews over the years changed a lot, in such a way that I am not any longer simply considering each point in the review, but am asking myself more often, what went wrong in the communication of the content.''} -- P13
\end{quoting}
Two participants (P15 and P24) reported doing \textit{informal discussion} of the reviews. Elaborating on their responses, participant P24 said that \textit{``[\ldots] Talking with peers who also submitted papers on review day also adds to stress if your paper wasn't accepted.''} On the other hand, participant P15 mentioned that going through the reviews with friends as a social event helps him/her to remain positive, especially when some comments in the reviews are disheartening. 


\section{Study Limitations}
The geographical diversity of the participants is limited to 10 countries from Europe and North America only. Moreover, 12 out of 24 (50\%) responses are from Germany and USA. We believe this might be the result of using our personal networks to circulate the questionnaire. Therefore, the analysis results may be dominated by the work culture of visualization PhD students in Europe and North America.

Since the questions in the third section of our questionnaire were open-ended and broad, we provided some examples for each question to specify along which lines participants can think of while answering the questions. On the one hand, those examples were good to encourage participants to share their thoughts in case they felt lost. But this might have influenced the responses to some extent on the other hand. 

Along similar lines, we might have biased the participants by advertising the study in a way that implied that they might feel frustrated about rejected papers---this was done to attract the attention of more participants. In general, just focusing on rejected papers takes a perspective that cuts short examples where papers were improved in a short revision cycle and then directly accepted. Since we had to compromise on the extensiveness of the questionnaire, we cannot make statements about this or other aspects of the submission process and strategies.


\section{Discussion}

Based on the responses, we can infer that PhD students in the field of visualization research take the reviews seriously. The students seem to discuss reviews and formulate a \textit{resubmission plan} and continue to make further \textit{improvements} in their work. The students frequently use the strategy of \textit{downgrading work} while resubmitting, either by shortening the length or opting for a lower-rated venue. It seems that the strategies work well as they eventually get their papers published. This also indicates that visualization research community has many venues and tracks where the improved work eventually gets published. The strategies reported by visualization PhD students---discussing reviews, improving the work, and resubmitting it to another venue---align well with the suggestions made by Shneiderman~\cite[Section 6.6]{shneiderman2016new} and Elmqvist~\cite{elmqvist2016dealing}.

Although the responses of PhD students indicate that reviews are painful at first, they try to \textit{see rejection as a positive} event. Some students mentioned that they try not to take the reviews personally and highlighted that rejections do not mean incompetency, which resonates with the advice by Elmqvist~\cite{elmqvist2018above}. Other students also mentioned that engaging in informal discussions with colleagues and friends helps them cope with paper rejection. Knowing about these strategies can be helpful especially for young PhD students who have limited knowledge about the normalcy of receiving rejections in a peer-review research process. It maybe a good idea to talk about rejections, both formally and informally, and discuss strategies on how to cope with them. The discussions potentially creates a social support system that helps in coping with rejections and matches well with the recommendation by Day~\cite{day2011thesilent}.
 
There were few responses which highlighted the occurrences of \textit{unclear} and \textit{contradictory reviews}. Even though these factors contribute to the painful experience on the day of notification, PhD students in visualization research usually handle them positively. As a result, many students highlighted that their impression of reviews gets better with time. Some students reported that they had used their judgment and sometimes had disagreed with few comments in the reviews.
Additionally, some experienced students mentioned doing a \textit{high-level reflection of reviews} trying to understand the issues that might have misdirected the reviewers. These strategies can also be helpful for finding areas of improvement and formulating a plan of action, especially when the reviews are unclear, have contradictory views, or contain insufficient information to back the review rating. The high-level reflection of reviews can be done in different ways. Elmqvist suggested summarizing the reviews~\cite{elmqvist2016dealing} while Shneiderman shared personal experiences of handling difficult reviews and highlighted using one's own judgment to continue improving the work~\cite[Section 6.6]{shneiderman2016new}. 


Since nine respondents (37.5\%) indicated to have experienced emotional distress on the day of rejection notification after reading the reviews, it may be indicative of a harsh textual connotation being used in the reviews. We argue that conscious efforts can be made by reviewers to proactively minimize the painful rejection experience. A more polite text tone over a harsh comment may be used to communicate the same review. Although not always possible, concrete suggestions for improving the work can be clearly highlighted. Moreover, alternative reviewing models can be explored and adopted that have helped in reducing the harsh tone; Besan{\c{c}}on \textit{et al.}~\cite{besanccon2020open} concluded in a recent survey that reviewers use a more polite tone in open and non-anonymized peer reviewing model.





\section{Conclusion and Future Work}

Based on the responses we can infer that most PhD students in the visualization research have a positive outlook to cope with paper rejections. However, sometimes the positive outlook does not correlate with the absence of a painful experience while facing rejections (\textit{e.g.}, P14). We hope that knowing about the strategies of others can help the students who are facing paper rejections. Also, we hope that initiating discussions on the normalcy of paper rejections in academia and talking about known strategies in coping with rejections can be helpful especially to the young PhD students joining a research team. 

As part of future work, we would be interested in discerning the strategies of students having different levels of experience. Another aspect would be to validate the results of this analysis by getting expert feedback of experienced reviewers from the community. Also, we plan to extend the analysis to the complete set of responses.

\section*{Acknowledgments}
We wish to thank all the researchers who participated in our study and shared their thoughts and experiences. We are indebted to Ben Shneiderman and Niklas Elmqvist who pointed us to valuable resources on the subject matter.


\bibliographystyle{abbrv-doi-hyperref}

\bibliography{template}

\begin{thebibliography}{10}

\bibitem{supplementary}
\href{https://doi.org/10.17605/osf.io/vgj36}{S.~Agarwal}.
\newblock \href{https://doi.org/10.17605/osf.io/vgj36}{Supplementary material:
  How visualization {PhD} students cope with paper rejections},
  \href{https://doi.org/10.17605/osf.io/vgj36}{2020}.
  \href{https://doi.org/10.17605/osf.io/vgj36}
{doi: {{%
10\hspace{.1pt}\discretionary{.}{%
}{.}\hspace{.4pt}17605\discretionary{/}{%
}{/}osf\hspace{.1pt}\discretionary{.}{%
}{.}\hspace{.4pt}io\discretionary{/}{%
}{/}vgj36}}}


\bibitem{besanccon2020open}
\href{https://doi.org/10.1186/s41073-020-00094-z}{L.~Besan{\c{c}}on,
  N.~R{\"o}nnberg, J.~L{\"o}wgren, J.~P. Tennant, and M.~Cooper}.
\newblock \href{https://doi.org/10.1186/s41073-020-00094-z}{Open up: a survey
  on open and non-anonymized peer reviewing}.
\newblock \href{https://doi.org/10.1186/s41073-020-00094-z}{{\em Research
  Integrity and Peer Review}},
  \href{https://doi.org/10.1186/s41073-020-00094-z}{5(1):1--11},
  \href{https://doi.org/10.1186/s41073-020-00094-z}{2020}.
  \href{https://doi.org/10.1186/s41073-020-00094-z}
{doi: {{%
10\hspace{.1pt}\discretionary{.}{%
}{.}\hspace{.4pt}1186\discretionary{/}{%
}{/}s41073\discretionary{%
}{-}{-}020\discretionary{%
}{-}{-}00094\discretionary{%
}{-}{-}z}}}


\bibitem{day2011thesilent}
\href{https://doi.org/10.5465/amle.2010.0027}{N.~E. Day}.
\newblock \href{https://doi.org/10.5465/amle.2010.0027}{The silent majority:
  Manuscript rejection and its impact on scholars}.
\newblock \href{https://doi.org/10.5465/amle.2010.0027}{{\em Academy of
  Management Learning \& Education}},
  \href{https://doi.org/10.5465/amle.2010.0027}{10(4):704--718},
  \href{https://doi.org/10.5465/amle.2010.0027}{2011}.
  \href{https://doi.org/10.5465/amle.2010.0027}
{doi: {{%
10\hspace{.1pt}\discretionary{.}{%
}{.}\hspace{.4pt}5465\discretionary{/}{%
}{/}amle\hspace{.1pt}\discretionary{.}{%
}{.}\hspace{.4pt}2010\hspace{.1pt}\discretionary{.}{%
}{.}\hspace{.4pt}0027}}}


\bibitem{edwards2018emotions}
\href{https://doi.org/10.1017/jmo.2018.20}{M.~S. Edwards and N.~M. Ashkanasy}.
\newblock \href{https://doi.org/10.1017/jmo.2018.20}{Emotions and failure in
  academic life: Normalising the experience and building resilience}.
\newblock \href{https://doi.org/10.1017/jmo.2018.20}{{\em Journal of Management
  \& Organization}},
  \href{https://doi.org/10.1017/jmo.2018.20}{24(2):167–188},
  \href{https://doi.org/10.1017/jmo.2018.20}{2018}.
  \href{https://doi.org/10.1017/jmo.2018.20}
{doi: {{%
10\hspace{.1pt}\discretionary{.}{%
}{.}\hspace{.4pt}1017\discretionary{/}{%
}{/}jmo\hspace{.1pt}\discretionary{.}{%
}{.}\hspace{.4pt}2018\hspace{.1pt}\discretionary{.}{%
}{.}\hspace{.4pt}20}}}


\bibitem{elmqvist2016dealing}
N.~Elmqvist.
\newblock Dealing with rejection.
\newblock
  \url{https://sites.umiacs.umd.edu/elm/2016/10/25/dealing-with-rejection/},
  2016.
\newblock accessed: July 2020.

\bibitem{elmqvist2018above}
N.~Elmqvist.
\newblock Above all, persistence.
\newblock
  \url{https://sites.umiacs.umd.edu/elm/2018/08/16/above-all-persistence/},
  2018.
\newblock accessed: July 2020.

\bibitem{sierk2016the}
\href{https://doi.org/10.1177/1056492615586597}{S.~A. Horn}.
\newblock \href{https://doi.org/10.1177/1056492615586597}{The social and
  psychological costs of peer review: Stress and coping with manuscript
  rejection}.
\newblock \href{https://doi.org/10.1177/1056492615586597}{{\em Journal of
  Management Inquiry}},
  \href{https://doi.org/10.1177/1056492615586597}{25(1):11--26},
  \href{https://doi.org/10.1177/1056492615586597}{2016}.
  \href{https://doi.org/10.1177/1056492615586597}
{doi: {{%
10\hspace{.1pt}\discretionary{.}{%
}{.}\hspace{.4pt}1177\discretionary{/}{%
}{/}1056492615586597}}}


\bibitem{munzner2008process}
\href{https://doi.org/10.1007/978-3-540-70956-5_6}{T.~Munzner}.
\newblock \href{https://doi.org/10.1007/978-3-540-70956-5_6}{Process and
  pitfalls in writing information visualization research papers}.
\newblock \href{https://doi.org/10.1007/978-3-540-70956-5_6}{In {\em
  Information Visualization: Human-Centered Issues and Perspectives}},
  \href{https://doi.org/10.1007/978-3-540-70956-5_6}{pp. 134--153}.
  \href{https://doi.org/10.1007/978-3-540-70956-5_6}{2008}.
  \href{https://doi.org/10.1007/978-3-540-70956-5_6}
{doi: {{%
10\hspace{.1pt}\discretionary{.}{%
}{.}\hspace{.4pt}1007\discretionary{/}{%
}{/}978\discretionary{%
}{-}{-}3\discretionary{%
}{-}{-}540\discretionary{%
}{-}{-}70956\discretionary{%
}{-}{-}5\_6}}}


\bibitem{rubin2014converting}
\href{https://doi.org/10.1201/b16720-55}{D.~B. Rubin}.
\newblock \href{https://doi.org/10.1201/b16720-55}{Converting rejections into
  positive stimuli}.
\newblock \href{https://doi.org/10.1201/b16720-55}{In {\em Past, Present, and
  Future of Statistical Science}}, \href{https://doi.org/10.1201/b16720-55}{pp.
  593--603}. \href{https://doi.org/10.1201/b16720-55}{2014}.
  \href{https://doi.org/10.1201/b16720-55}
{doi: {{%
10\hspace{.1pt}\discretionary{.}{%
}{.}\hspace{.4pt}1201\discretionary{/}{%
}{/}b16720\discretionary{%
}{-}{-}55}}}


\bibitem{shneiderman2016new}
B.~Shneiderman.
\newblock {\em The new ABCs of research: Achieving breakthrough
  collaborations}.
\newblock 1st ed., 2016.

\end{thebibliography}
\end{document}